\documentclass[12pt]{article}
\usepackage{a4,epsfig}
\begin{document}

\begin{center}
{\bf \large  A Systematic Investigation of Light Heavy-Ion
Reactions
\\}
I. Boztosun \footnote{Present address: Department of Nuclear Physics,
University of Oxford, Keble Road OX1 3RH Oxford UK} \\
Department of Physics, Erciyes University, 38039 Kayseri Turkey \\
E-mail: i.boztosun1@physics.ox.ac.uk
\end{center}

\begin{abstract}
We introduce a novel coupling potential for the scattering of
deformed light heavy-ion reactions. This new approach is based on
replacing the usual first-derivative coupling potential by a new,
second derivative coupling potential in the coupled-channels
formalism. The new approach has been successfully applied to the
study of the $^{12}$C+$^{12}$C, $^{12}$C+$^{24}$Mg,
$^{16}$O+$^{28}$Si and $^{16}$O+$^{24}$Mg systems and made major
improvements over all the previous coupled-channels calculations
for these systems. This paper also shows the limitations of the
standard coupled-channels theory and presents a global solution to
the problems faced in the previous theoretical accounts of these
reactions.
\end{abstract}

\section{Introduction}
We investigate the elastic and inelastic scattering of light
heavy-ions, which have stimulated a great deal of interest over
the last 40 years. There has been extensive experimental effort to
measure the elastic and inelastic scattering data as well as their
90$^{\circ}$ and 180$^{\circ}$ excitation functions. A large body
of experimental data for these systems is available (see
\cite{Bra82,Sci97,Sto79,Kob84} and references therein). A variety
of theoretical accounts based on dynamical models or purely
phenomenological treatments have been proposed to explain these
data \cite{Bra82,Bra97}. The elastic scattering data have already
been studied in detail using optical model and coupled-channels
method.

Although most of these models provide reasonably good fits, no
unique model has been proposed that explains consistently the
elastic and inelastic scattering data over a wide energy range
without applying any {\it ad hoc} approaches. Consequently, the
following problems continue to exist for light heavy-ion
reactions: $(1)$ explanation of anomalous large angle scattering
data, `ALAS'; $(2)$ reproduction of the oscillatory structure near
the Coulomb barrier; $(3)$ the lack of the correct oscillatory
structure between theoretical predictions and experimental data
for the ground and excited states; $(4)$ simultaneous fits of the
individual angular distributions, resonances and excitation
functions (for the $^{12}$C + $^{12}$C system in particular);
$(5)$ the magnitude of the mutual-2$^{+}$ excited state data in
the $^{12}$C + $^{12}$C system is unaccounted for; $(6)$ the
deformation parameters ($\beta$ values): previous calculations
require $\beta$ values that are at variance with the empirical
values and are physically unjustifiable.

Therefore, in this paper, we are concerned with the measured
experimental data for $^{12}$C+$^{12}$C, $^{16}$O+$^{28}$Si,
$^{12}$C+$^{24}$Mg and $^{16}$O+$^{24}$Mg in an  attempt to find a
global model, which simultaneously fits the elastic and inelastic
scattering data for the ground and excited states in a consistent
way over a wide energy range, and which throws light on the
underlying mechanism of the reactions and on the nature of the
interactions involved.

\section{Standard Coupled-Channels Model}

Although we have considered and studied four different reactions,
we will show some of the results for the $^{16}$O+$^{28}$Si and
$^{12}$C+$^{12}$C reactions. The details of the models and a
complete set of the results for all reactions can be found in
references \cite{Boz1,Boz2,Boz3,Boz4}.

We describe the interaction between $^{16}$O and $^{28}$Si nuclei
with a deformed optical potential. The real potential is assumed
to have the square of a Woods-Saxon shape:
\begin{equation}
V_{N}(r) = \frac{-V_{0}}{(1+exp(r-R)/a)^{2}}
\end{equation}
with $V_{0}$=706.5 MeV, R=$r_{0}$($A_{p}$$^{1/3}$+$A_{t}$$^{1/3}$)
with $r_{0}$=0.7490 fm  and a=1.40 fm. The parameters of the real
potential were fixed as a function of energy and were not changed
in the present calculations although it was observed that small
changes could improve the quality of the fits.

The imaginary part of the potential was taken as in
ref.\cite{Kob84} as the sum of a Woods-Saxon volume and surface
potential:
\begin{equation}
W(r)=-W_{V}f_{(r,R_{V},a_{V})}+4W_{S}a_{S}df_{(r,R_{S},a_{S})}/dr
\end{equation}
\begin{equation}
f_{(r,R,a)} = \frac{1}{(1+exp((r-R)/a))}
\end{equation}
with $W_{V}$=59.9 MeV, $a_{V}$=0.127 fm and $W_{S}$=25.0 MeV,
$a_{S}$=0.257 fm. These parameters were also fixed in the calculations
and only their radii increased linearly with energy according to the
following formulae.
\begin{equation}
R_{V} = 0.06084E_{CM}-0.442
\end{equation}
\begin{equation}
R_{S} = 0.2406E_{CM}-2.191
\end{equation}

Since the target nucleus $^{28}$Si is strongly
deformed, it is essential to treat its collective
excitation explicitly in the framework of the coupled-channels
formalism. It has been assumed that the target nucleus has a static
quadrupole deformation, and that its rotation can be described in
the framework of the collective rotational model. It is therefore
taken into account by deforming the real optical potential in the following
way
\begin{equation}
R(\theta,\phi)=r_{0}A_{p}^{1/3}+r_{0}A_{t}^{1/3}[1+\beta_{2}
Y_{20}(\theta,\phi)]
\end{equation}
where $\beta_{2}$=-0.64 is the deformation parameter of $^{28}$Si.
This value is actually larger than the value calculated from the
known BE(2) value. However this larger $\beta_{2}$ was needed to
be able to fit the magnitude for the 2$^{+}$ data.

In the present calculations, the first two excited states of the
target nucleus $^{28}$Si: 2$^{+}$ (1.78 MeV) and 4$^{+}$ (4.62
MeV) were included and the 0$^{+}$-2$^{+}$-4$^{+}$ coupling scheme
was employed. The reorientation effects for 2$^{+}$ and 4$^{+}$
excited states were also included. The calculations were performed
with an extensively modified version of the code
CHUCK~\cite{Kunz}.

Using the standard coupled-channels theory, we found, as other
authors had found, that it was impossible to describe consistently
the elastic and inelastic scattering of this and other reactions
we considered.

\section{New Coupling Potential}
\begin{figure}
\centering \epsfxsize=8cm \hskip+0.75cm \epsfbox{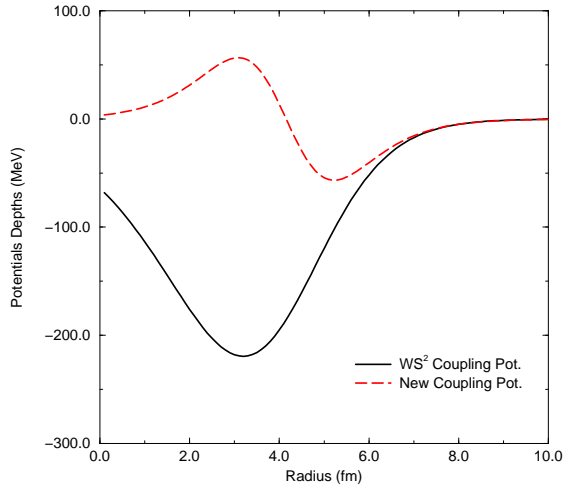}
\vskip-4.5cm \caption{For $^{16}$O+$^{28}$Si, the comparison of
the {\it standard coupling potential} which is the first
derivative of the central potential and our {\it new coupling
potential} which is parameterised as the 2$^{nd}$ derivative of
Woods-Saxon shape and which has V=155.0 MeV, R=4.160 fm and a=0.81
fm.} \label{16Ocoup}
\end{figure}

The limitations of the standard coupled-channels theory in the
analyses of these reactions compelled us to look for another
solution. Therefore, a second-derivative coupling potential, as
shown in figure \ref{16Ocoup}, has been used in the place of the
usual first-derivative coupling potential. The interpretation of
this new coupling potential is given in reference \cite{Boz5}. We
employed the same method with small changes in the potential
parameters. The empirical deformation parameter ($\beta_{2}$) is
used in these new calculations.
\section{Results}
\subsection{$^{16}$O+$^{28}$S\lowercase{i}}
The first system we consider is $^{16}$O + $^{28}$Si, which shows
anomalous large angle scattering (ALAS). In the present work, we
consider an extensive {\it simultaneous} investigation of the
elastic and the inelastic scattering of this system at numerous
energies from $E_{Lab}$=29.0 MeV to 142.5 MeV over the whole
angular range up to 180$^{\circ}$. In this energy range, the
excitation functions for the ground and 2$^{+}$ states are also
analysed \cite{Boz1,Boz6}.

Several {\it ad hoc} models have been proposed to explain these
data, but no satisfactory microscopic models have been put forward
yet. The most satisfactory explanation proposed so far is that of
Kobos and Satchler \cite{Kob84} who attempted to fit the elastic
scattering data with a microscopic double-folding potential.
However, these authors had to use some small additional {\it ad
hoc} potentials to obtain good agreement with the experimental
data.

\begin{figure}
\epsfxsize=12.5cm \centering \epsfbox{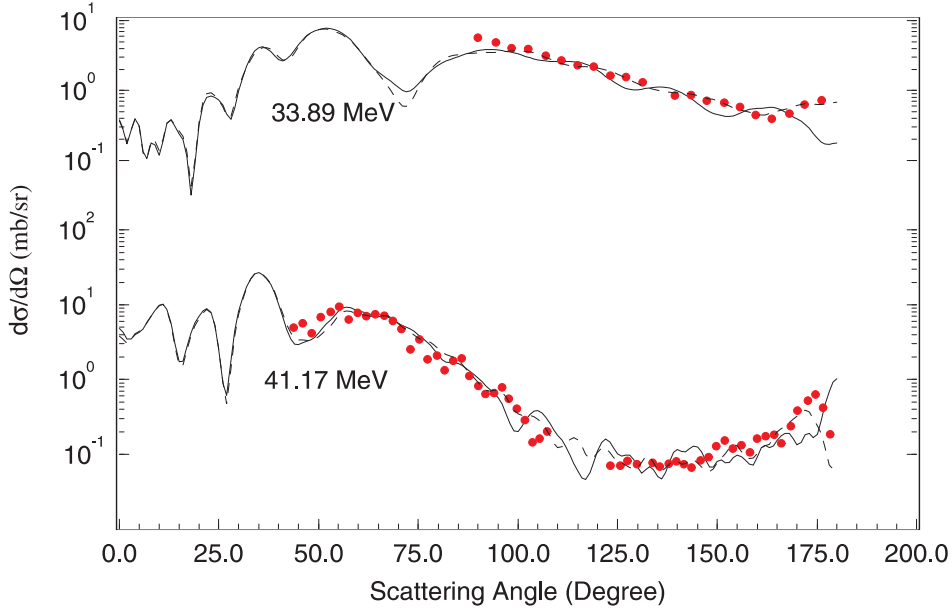} \vskip-0.0cm
\caption{$^{16}$O+$^{28}$Si system: The {\it solid} lines are the
results of {\it standard} coupled-channels and {\it dashed} lines
are the results obtained using new coupling potential for the
inelastic scattering data.} \label{16Oexccomp}
\end{figure}

Using the standard coupled-channels method, some of the results
obtained for the 180$^{\circ}$ excitation functions for the ground
and 2$^{+}$ states of the $^{16}$O+$^{28}$Si reaction are shown in
figure \ref{16Oexccomp}. The magnitude of the cross-sections and
the phase of the oscillations for the individual angular
distributions are given correctly at most angles. However, there
is an out-of-phase problem between the theoretical predictions and
the experimental data towards large angles at higher energies.
This problem is clearly seen in the 180$^{\circ}$ excitation
functions which are shown in the figure. A number of models have
been proposed, ranging from isolated resonances to cluster
exchange between the projectile and target nucleus to solve these
problems (see ref.\cite{Bra82} for a detailed discussion).

We have also attempted to overcome these problems by considering:
$(i)$ changes in the real and imaginary potentials, $(ii)$
inclusion of 6$^{+}$ excited state, $(iii)$ changes in the
$\beta_{2}$ value, $(iv)$ the inclusion of the hexadecapole
deformation ($\beta_{4}$). These attempts failed to solve the
problems at all \cite{Boz1,Boz6}. We were unable to get an
agreement with the elastic and the 2$^{+}$ inelastic data as well
as the 180$^{\circ}$ excitation functions simultaneously within
the standard coupled-channels formalism.
However, as shown in figure~\ref{16Oexccomp}, the new coupling potential has solved
the out-of-phase problem for the 180$^{\circ}$ excitation functions and fits the
ground state and 2$^{+}$ state data simultaneously.
\subsection{$^{12}$C+$^{24}$M\lowercase{g} \lowercase{and} $^{16}$O+$^{24}$M\lowercase{g}}
The second and third examples we have considered are
$^{12}$C+$^{24}$Mg and $^{16}$O+$^{24}$Mg. Fifteen complete
angular distributions of the elastic scattering of
$^{12}$C+$^{24}$Mg system were measured at energies around the
Coulomb barrier and were published recently~\cite{Sci97}. We have
studied these fifteen complete elastic scattering angular
distributions as well as some inelastic scattering data measured
by Carter et al. some twenty years ago~\cite{carter1,carter2}.
Excellent agreement with the experimental data was obtained by
using this new coupling potential. Our model has also solved some
problems in $^{16}$O+$^{24}$Mg scattering \cite{Boz3}.

\subsection{$^{12}$C+$^{12}$C}
\begin{figure}[t]
\epsfxsize=8.5cm  \centering \epsfbox{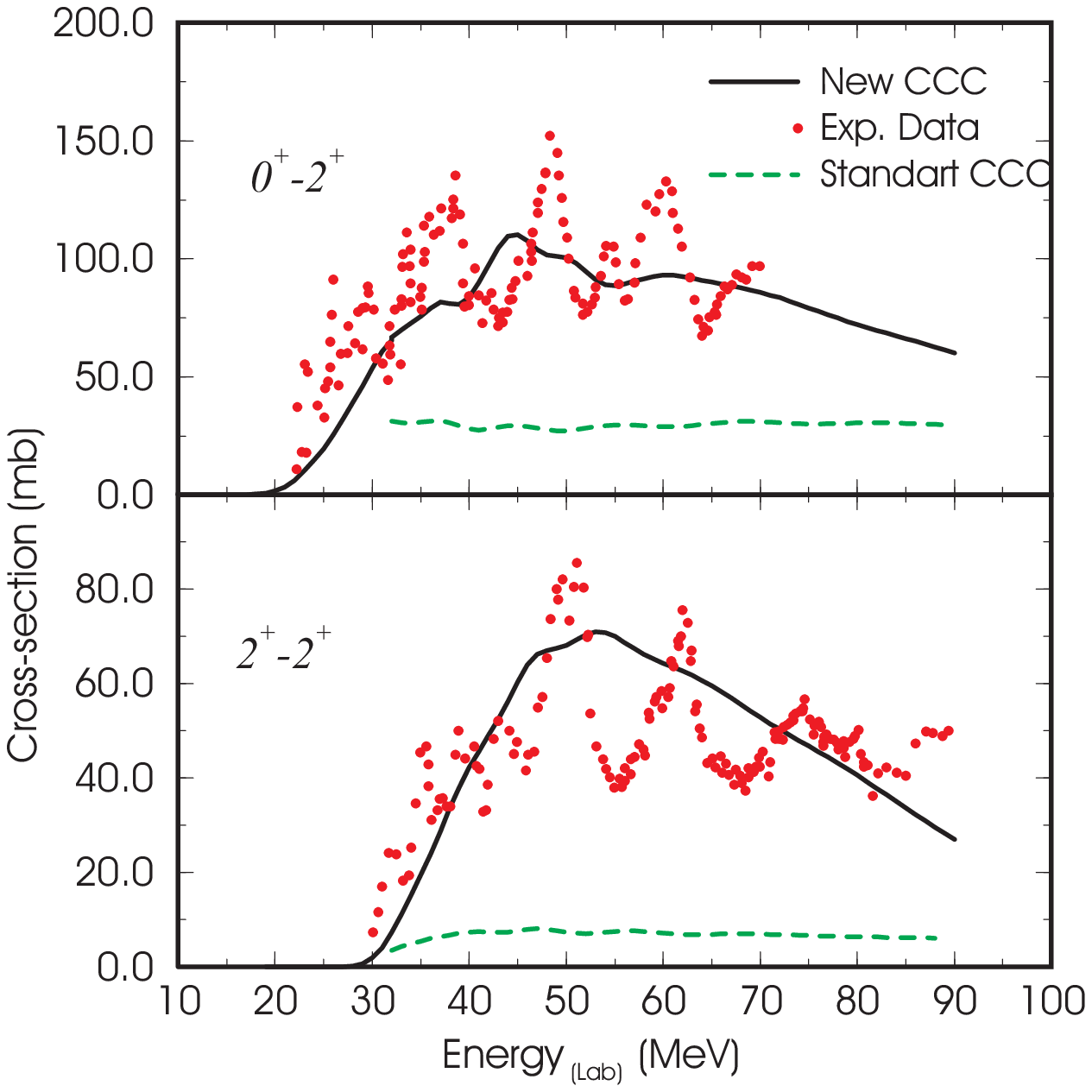} \vskip-0.0cm
\caption{$^{12}$C+$^{12}$C system: The integrated cross-section of
the single and mutual-2$^{+}$ states. The {\it solid} lines are
the results of {\it the new coupling potential}, while the {\it
dashed} lines are the results of {\it standard coupled-channels
model}.} \label{comp}
\end{figure}

\begin{figure}
\epsfxsize=11.5cm \centering \hskip+1.5cm \epsfbox{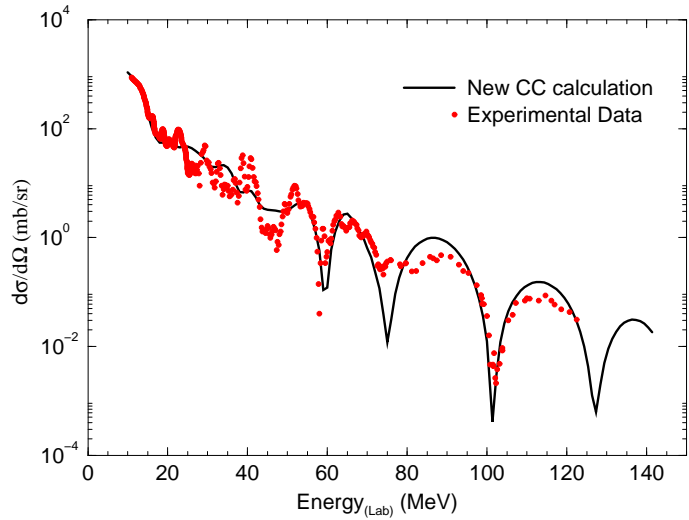}
\vskip-9.0cm \caption{$^{12}$C+$^{12}$C system: 90$^{\circ}$
excitation function for the elastic scattering using {\it new}
coupling potential.} \label{12Cexc}
\end{figure}

The final system we have considered is that of $^{12}$C+$^{12}$C,
which has been studied extensively over the last 40 years. There
has been so far no model, which fits consistently the elastic,
inelastic scattering data and mutual excited state data as well as
the resonances and excitation functions. Another problem is the
predicted magnitude of the excited state cross-sections, in
particular for the mutual-2$^{+}$ channel. The conventional
coupled-channels model underestimates its magnitude by a factor of
at least two and often much more~\cite{wolf,sakuragi,Fry}. We have
also observed this in our conventional coupled-channels
calculations as shown in figure~\ref{comp} with dashed lines.
There are also resonances observed at low energies, which have
never been fitted by a potential, which also fits either the
angular distributions or the excitation functions. Therefore, the
experimental data at many energies between 20.0 and 126.7 MeV in
the laboratory system have been studied simultaneously to attempt
to find a global potential.
Using this new coupling potential, we have been able to fit the
energy average of all the available ground, single-2$^{+}$,
mutual-2$^{+}$ and the backgrounds in the integrated
cross-sections as well as the main gross features of the
90$^{\circ}$ excitation function, as shown in figures \ref{comp}
and \ref{12Cexc}, simultaneously. Our preliminary calculations of
resonances using no imaginary potential are promising but there
are problems with the widths of the resonances.

To summarise while these four systems show quite different
properties and problems, a unique solution has come from a new
coupling potential. Although the origin of this new coupling
potential is still speculative and needs to be understood from a
microscopic viewpoint, the approach outlined here is universal and
applicable to all the systems. Studies using this new coupling
potential are likely to lead to new insights into the formalism
and the interpretation of these systems. Therefore, this work
represents an important step towards understanding the elastic and
inelastic scattering of light deformed heavy-ion systems.

\section*{Acknowledgments}
Special thanks to W.D.M Rae, Y. Nedjadi, S. Ait-tahar, G.R.
Satchler and D.M. Brink for valuable discussions and providing
some data. I. Boztosun also would like to thank the Turkish
Council of Higher Education (Y\"{O}K), Oxford and Erciyes
Universities, Turkey, for their financial support.

\end{document}